\documentstyle[emulateapj,apjfonts,psfig]{article}

\lefthead{Miller, Cackett, \& Reis}  
\righthead{Absorption in X-ray Binaries}

\received{Date}
\revised{Date}
\accepted{Date}

\journalid{vol}{date}
\articleid{1}{4}
\paperid{id}

\cpright{AAS}{1999}
\ccc{x}

\begin{document}

\title{On Neutral Absorption and Spectral Evolution in X-ray Binaries}

\author{J.~M.~Miller\altaffilmark{1},
        E.~M.~Cackett\altaffilmark{1,2},
     	R.~C.~Reis\altaffilmark{3}}
       
\altaffiltext{1}{Department of Astronomy, University of Michigan, 500
Church Street, Ann Arbor, MI 48109-1042, jonmm@umich.edu}

\altaffiltext{2}{Chandra Fellow}

\altaffiltext{3}{Institute of Astronomy, University of Cambridge,
Madingley Road, Cambridge CB3 OHA, UK}


\authoremail{jonmm@umich.edu}

\label{firstpage}

\begin{abstract}
Current X-ray observatories make it possible to follow the evolution
of transient and variable X-ray binaries across a broad range in
luminosity and source behavior.  In such studies, it can be unclear
whether evolution in the low energy portion of the spectrum should be
attributed to evolution in the source, or instead to evolution in
neutral photoelectric absorption.  Dispersive spectrometers make it
possible to address this problem.  We have analyzed a small but
diverse set of X-ray binaries observed with the {\it Chandra} High
Energy Transmission Grating Spectrometer across a range in luminosity
and different spectral states.  The column density in individual
photoelectric absorption edges remains constant with luminosity, both
within and across source spectral states.  This finding suggests that
absorption in the interstellar medium strongly dominates the neutral
column density observed in spectra of X-ray binaries.  Consequently,
evolution in the low energy spectrum of X-ray binaries should properly
be attributed to evolution in the source spectrum.  We discuss our
results in the context of X-ray binary spectroscopy with current and
future X-ray missions.
\end{abstract}

\section{Introduction}
X-ray binaries show a remarkable variety of extreme phenomena,
including rapid variability in many wavelength bands, the
production of relativistic jets and hot winds, and outbursts that span
factors of $10^{6}$ in luminosity (for a review of black hole
transients, see Remillard \& McClintock 2006).  Monitoring
observations with the {\it Rossi X-ray Timing Explorer} and {\it
Swift} enable the evolution of X-ray spectral and timing properties
of X-ray binaries to be traced in order to understand the accretion
process (see, e.g., Rykoff et al.\ 2007).  Model degeneracies that may
occur in fitting a single spectrum can often be resolved by
examining how multiple spectra evolve with time.

Major changes within the outbursts of transient and persistent X-ray
binaries are typically linked with changes in the nature of the
accretion disk (see, e.g., Esin, McClintock, \& Narayan 1997; Homan et
al.\ 2001; also see Miller et al.\
2006a).  Tracking accretion disk evolution is therefore a primary goal
of monitoring observations.  The disk is directly visible in soft
X-rays in black hole X-ray binaries, and likely also in neutron star
binaries (e.g. Cackett et al.\ 2008a).  However, the K shell
photoelectric absorption edges from most abundant elements fall below
2~keV, as do Fe L shell edges.  This absorption can complicate efforts
to study disk properties, especially when using a spectrometer with
limited energy range and resolution.

Prominent theoretical treatments of accretion flows predict modest
winds (e.g. Begelman, McKee, \& Shields 1983) and/or outflows with a
high ionization parameter (e.g. Narayan \& Raymond 1999).
Observationally, even the most extreme winds detected in X-ray
binaries do not appear to generate enough column density to add
appreciably to the neutral absorption column (Miller et al.\ 2006).
Nevertheless, it is not uncommon to allow the equivalent neutral
hydrogen column density to vary in modeling multiple spectra of an
X-ray binary (see, e.g. Brocksopp et al.\ 2006; Caballero-Garcia et
al.\ 2009; Cabanac et al.\ 2009).  Even in cases where the neutral
column density is held constant or jointly determined, this method is
adopted somewhat arbitrarily, without specific reference to an
observational or theoretical basis (see, e.g., Sobczak et al.\ 2000;
Miller et al.\ 2001; Nowak, Wilms, \& Dove 2002; Park et al.\ 2004;
Gierlinski \& Done 2004; Kalemci et al.\ 2005; Rykoff et al.\ 2007).

Dispersive X-ray spectroscopy can separate variations in absorption
from variations in the source spectrum.  Even the resolution afforded
by CCD spectrometers blends many weak edges into the continuum, and
the optical depth and wavelength of stronger edges can be difficult to
measure precisely.  For instance, in fits to CCD spectra, the multiple
Fe L edges can be modeled satisfactorily in terms of a single step
function.  In constrast, {\it Chandra} High Energy Transmission
Grating Spectrometer observations of X-ray binaries have resolved the
individual Fe L edges and their detailed properties (see, e.g., Schulz
et al. 2002, Juett et al.\ 2006).  Beyond simple edges, dispersive
X-ray spectroscopy has clearly revealed the neutral O I 1s-2p
resonance absorption line that was first glimpsed with the crystal
spectrometer aboard {\it Einstein} (Schattenberg \& Canizares 1986,
Paerels et al.\ 2001).

In this Letter, we extend recent work using high resolution X-ray
spectroscopy by examining the evolution of neutral absorption in a
sample of X-ray binaries.  We find no evidence for variable
absorption, consistent with an interstellar origin for neutral absorption.

\section{Data Selection and Reduction}
Our analysis is restricted to observations made with the {\it Chandra}
High Energy Transmission Grating Spectrometer.  The HETGS offers
better resolution than the {\it XMM-Newton} Reflection Grating
Spectrometer, as well as coverage across the full 0.3-10.0~keV band.
When the equivalent neutral hyrdogen density is too high, much of the
source continuum spectrum is scattered out of the line of sight.  This must be
avoided as absorption features are measured relative to a continuum.
A number of clear absorption edges are detected in Cygnus X-1, which
has an equivalent neutral hydrogen column density of $N_{H} =
6.2\times 10^{21}~ {\rm cm}^{-2}$ (Schulz et al.\ 2002).  However, an
estimated column density of $8.0\times 10^{21}~ {\rm cm}^{-2}$ does
not allow the absorption edges in XTE J1550$-$564 to be studied in
detail (Miller et al.\ 2003).  We therefore restricted our analysis to
sources equal to or below approximately $6\times 10^{21}~ {\rm
cm}^{-2}$.  

Our analysis was also restricted by the need to select variable and
transient sources. Observations of transient sources with
large missions can be difficult, and relatively few sources with low
or moderate column densities have been observed on multiple occasions.
The sources and observations selected for our analysis are listed in
Table 1.  Different source types, companion types,
and orbital periods are included in the sample.

When observing bright sources using the HETGS, the ACIS-S array of
CCDs must be run in ``continuous clocking'' mode to prevent photon
pile-up.  For consistency, the observations considered in this
work employed this observing mode.  The only exception is a single
observation of GX~339$-$4 in a low flux state (obsid 4420; see Table
1).  An observation of Cygnus X-1 in the low/hard state that was not
taken in ``continuous clocking'' mode (obsid 107) suffered heavy
photon pile-up, complicating absorption spectroscopy (see Juett,
Schulz, \& Chakrabarty 2004), and was therefore rejected.  For more
information on the specific nature of HETGS observations made in
``continuous clocking'' mode, please see Miller et al.\ (2006b) or
Cackett et al.\ (2008b).

Most of the spectra considered in this work were obtained using
the ``TGCat'' data center (see http://tgcat.mit.edu).  This facility is
run by the {\it Chandra} X-ray Center and provides first-order HETGS
spectra and responses derived using up-to-date CIAO software and
calibrations.  An error in the instrument offsets used to observe XTE
J1817$-$330 required custom processing as per the procedure outlined
in Miller et al.\ (2006b) and Cackett et al.\ (2008b).  Two
observations of 4U 1820$-$30 were previously reduced and analyzed by
Cackett et al.\ (2008b) (obsid 6633 and 6634), and the same spectra are
used in this work.  Using CIAO version 4.0.2, we combined the
first-order MEG spectra and instrument response files using the CIAO
tool ``add\_gratings\_spectra''.  The same procedure was repeated for
the first-order HEG spectra.  Again owing to the pointing error, only
the minus-side first-order spectra of XTE J1817$-$330 were used in our
analysis.  To create distinct spectra of Cygnus X-2 in the flaring,
horizontal, and normal branches of its ``Z'' track, we employed
exactly the same selection criteria and procedure as Schulz et al.\
(2009) to both observations (obtaining a total of six spectra).  The
``aglc'' script was used in the ISIS analysis package (Houck \&
Denicola 2000) to create lightcurves and color-color diagrams that
were used to create spectra from each branch.

All of the spectral fits reported in this work were made using XSPEC
version 12 (Arnaud \& Dorman 2000).  Unless otherwise stated, errors
are $1\sigma$ errors.

\section{Analysis and Results}
The true neutral absorption column density observed in a given
spectrum should not be a function of the continuum flux model.
However, to ensure model-independent results, we fit for prominent neutral
photoelectric absorption edges 

\centerline{~\psfig{file=f1.ps,width=3.1in,angle=-90}~}
\figcaption[t]{\footnotesize In the plot above, high/soft (black) and
low/hard (red) spectra of Cygnus X-1 in the region of the Ne K edge
are plotted as a ratio to a power-law continuum.  The best-fit column
density for the edge has been set to zero to show the remarkable
consistency of the edge in different spectral states.}
\medskip

\noindent individually -- in narrow wavelength
ranges -- using local power-law models for the continuum.  The O K, Fe
L, and Ne K edges were fit in the 20--25\AA, 16--18\AA, and 13--15\AA~
bands, respectively.  At longer wavelengths, the MEG has a higher
effective area than the HEG; moreover, it covers longer wavelengths
than the HEG.  Fits to absorption edges were therefore made to the
combined first-order MEG spectrum from each observation.

In many of the spectra, it was quickly apparent that simple
step-function edge models do not fully describe the data.  We
therefore used the ``tbnew'' model (Wilms et al.\ 2009), which
includes up-to-date cross-sections and detailed edge structure.  A new
functionality in ``tbnew'' can be used to fit the column density in an
individual edge by setting the overall column density and abundance of
all elements (apart from the element of interest) to zero, and
restricting the column density to negative values for the parameter of
interest.  (Negative values instruct the model to return the column
density for the element; positive values return the abundance of a
given element relative to hydrogen.)  All fitting parameters relating
to grain properties were fixed at their default value.  In our fits,
then, ``tbnew'' functioned as a model with only one variable parameter
(the column density of the element of interest).  Due to a combination
of a relatively high total absorption column and the particular choice
of instrumental pointing offsets used, useful constraints on the O K
edge could not be obtained from existing data on GX~339$-$4, 4U
1820$-$30, and Cygnus X-2.  We note that Yao et al.\ (2009) were able
to study the Oxygen edge in Cygnus X-2 by adding multiple spectra, but
this procedure is inconsistent with the aim of this study.

To estimate the luminosity of each source in a given observation, we
fit the first-order HEG spectra in the 1.2--10.0~keV range using
simple absorbed continuum models.  The equivalent neutral hydrogen
column density was allowed to vary but values were found to be broadly
consistent with fits to individual edges.  These models should be
regarded as fiducial.  Some of the fits obtained are not formally
acceptable, owing to calibration uncertainties and/or unrelated
ionized absorption lines in the spectra.  The black hole sources were
fit with a continuum consisting of a disk blackbody and a power-law.
The neutron star spectra were all fit with a continuum consisting of a
simple blackbody function and a power-law.  To represent the evolution
of absorption with luminosity, the luminosity in each observation was
calculated for a given distance.  Errors on the luminosity were then
calculated using the $1\sigma$ errors on the continuum flux.  The
distances assumed in order to calculate luminosity values are listed
in Table 1.

Some of the spectra modeled in this paper have not been analyzed
previously.  All four {\it Chandra} observations of XTE J1817$-$330
were made in the high/soft state.  This is evident from inner disk
color temperatures that range between kT$=$0.48~keV and 0.94~keV, and
power-law indices that range between $\Gamma=2.4$ and 4.0.  The
evolution of the bright phase of the outburst of XTE J1817$-$330 is
detailed in Rykoff et al.\ (2007), and comparisons to that work again
indicates that the {\it Chandra} observations were all made in
high/soft states.  Similarly, spectral results have not been reported
based on fits to the second and third observations of GX 339$-$4 that
are considered in this work (obsids 4569 and 4570).  Here again, the
flux was strongly dominated by the accretion disk, with inner disk
color temperatures of $kT = 0.68$~keV and 0.81~keV.  Finally, the
third observation of 4U 1820$-$30 (obsid 7032) has not been published
previously, but analysis as per Cackett et al.\ (2008b, 2009) suggests
that the source was observed in the ``banana branch''.

We find that neutral absorption is remarkably constant with source
luminosity, and consistent across different spectral states.  Figure 1
shows the Ne K edge in the high/soft and low/hard state spectra of
Cygnus X-1.  The jump in the continuum at the Ne K edge is consistent.
Figure 2 plots the column density measured in individual absorption
edges versus luminosity for the sources in our sample.  For the black
hole sources, the column density in a given edge is constant with
luminosity within $1\sigma$ errors.  The case is much the same with
the neutron star X-ray binaries.  The column density measured in the
Ne K edge of 4U 1820$-$30 is marginally different at the $1\sigma$
level in two observations, but both are consistent with a third
observation.  All three columns agree at the 90\% confidence level.
The same holds true for two observations of Cygnus X-2.  

The Ne column densities that we have measured agree with values
reported by Juett, Schulz, \& Chakrabarty (2004) for 4U 1820$-$30,
Cygnus X-2, and GX 339$-$4.  The Ne column density that we measured in
Cygnus X-2 also agrees with a detailed analysis made by Yao et al.\
(2009).  Our value of the Ne column density agrees with that measured
in Cygnus X-1 by Hanke et al.\ (2009) though our values of the O and
Fe columns differ by factor of approximately two.  These disparities
can likely the result of different continuum modeling procedures and
implementations of ``tbnew''.

\section{Discussion and Conclusions}
To better understand the evolution of the low energy spectrum of X-ray
binaries, we made fits to individual photoelectric absorption edges in
high resolution X-ray spectra of selected sources.  The column density
measured in individual edges is not observed to vary across
different spectral states, nor over a broad range in luminosity (see
Table 1 and Figure 2).  This suggests that gas from X-ray binaries
is not typically an important source of the neutral absorption observed
in the spectra of these systems.  Rather, neutral absorption must be
dominated by the ISM.  A similar conclusion was reached by Juett,
Schulz, \& Chakrabarty (2004) based on upper limits on the velocity
dispersion of the ISM as measured through absorption lines.  Evolution
in the low energy spectrum of typical X-ray binaries, then, is best
attributed to evolution in the source continuum.

Neutral absorption in X-ray spectra is often fit by a single model
that parameterizes the accumulated absorption from individual edges as
an equivalent neutral hydrogen column density.  Values obtained from
high resolution spectra are likely to give the best measure of a true
equivalent total column density.  In practice, instrumental problems
such as internal scattering, carbon build-up from optical blocking
filters, and gain drift could prevent the adoption of a
gratings-derived value for the neutral column.  In such cases, our results
suggest that a single value of the equivalent neutral hydrogen column
density should be used to fit multiple spectra from monitoring
observations of a given source with a given detector.

The outflows that are observed in X-ray binaries are highly
ionized -- dominated by He-like and H-like charge states (see, e.g.,
Lee et al.\ 2002, Miller et al.\ 2004, Miller et al.\ 2006b, Schulz et
al.\ 2008).  An especially dense wind was observed in GRO
J1655$-$40, and even in that case the ionized columns observed are
insufficient to create strong absorption edges that could be mistaken
for additional neutral absorption (Miller et al.\ 2006c, 2008).
Moreover, in sources such as H 1743$-$322, GRO J1655$-$40, and GRS
1915$+$105, a paradigm is emerging wherein ionized winds are active in
soft, disk-dominated states, but absent in spectrally hard states like
those that typically hold when sources accrete at a low fraction of
their Eddington limit (Miller et al.\ 2006b, 2006c,
Miller et al.\ 2008, Neilsen \& Lee 2009).  In
Figure 2, it is clear that any variation in ionized winds
across states does not affect the column density measured in neutral
edges.

Our results are based on spectra which only reach down to
approximately $0.01~{\rm L}_{\rm Edd.}$.  Sensitive high-resolution X-ray
spectra that would permit strong constraints on absorption variability
have not yet been obtained from sources at lower luminosity.  However,
theoretical arguments again point to ionized outflows that would
contribute little to a neutral column.  Winds from advection-dominated
accretion flows are expected to be extremely hot (since advective
flows are very hot), and line spectra should be dominated by He-like
and H-like charge states (e.g. Narayan \& Raymond 1999).  Recent
observations of the stellar-mass black hole V404 Cyg, which accretes
at about $10^{-5}~ {\rm L}_{\rm Edd.}$, are able to rule-out the
winds predicted by some advective models (Bradley et al.\ 2007).  This
further suggests that ionized outflows are not likely to contribute
significantly to line-of-sight absorption, even as sources approach
quiescence.  

There are at least two classes of sources where our results may not
hold in all circumstances.  Neutron star X-ray binaries known as ``dippers''
are viewed at high inclinations, and material in the outer disk can
block emission from the central engine (see, e.g., Diaz Trigo et al.\
2006).  Within such flux dips, the observed neutral absorption column
may vary due to the changing geometry within the binary.  Similarly,
winds from massive stars are known to be clumpy and to sometimes cause
flux dips; some of these dips may also cause variations in the
observed equivalent neutral hydrogen column density
(e.g. Balucinska-Church et al.\ 2000).

\vspace{0.1in}

\noindent We thank Joern Wilms for creating the ``tbnew'' model used
in this work, and for guidance on its implementation.  We acknowledge
helpful comments from the anonymous referee that improved this paper.
J.M.M. gratefully acknowledges funding from the {\it Chandra} Guest
Observer program.  EMC gratefully acknowledges support provided by
NASA through the Chandra Fellowship Program, grant number PF8-90052.
RCR acknowledges STFC for financial support.

\begin{table}[t]
\begin{center}
\vspace{0.1in}
\begin{footnotesize}
\begin{tabular}{llllllll}
Source Name  &  ${\rm P}_{\rm orb}$ (days) &  distance (kpc) & Companion & ID   &     Start (MJD) &  Exposure (ksec) & State \\
\hline
\hline
GX 339$-$4   & 1.76$^{a}$  & 8.5$^b$ & low-mass$^{a}$ & 4420  &    52715.8   &    76.2 & hard/intm.$^{k}$  \\
     ~       & ~  & ~  & ~ & 4569  &    53239.5   &    50.1 & high/soft$^{l}$ \\
     ~       & ~  & ~  & ~ & 4570  &    53282.    &    50.2 & high/soft$^{l}$ \\
\hline
XTE J1817$-$330 & short$^{c}$  &  $8.5^{d}$  & low-mass$^{c}$ & 6615  &    53779.1   &    50.0 & high/soft$^{l}$ \\
         ~      & ~  & ~ & ~ & 6616  &    53790.2  &     49.6 & high/soft$^{l}$  \\
         ~      & ~  & ~ &  ~ & 6617  &    53809.3  &     46.7 & high/soft$^{l}$ \\
         ~      & ~  & ~  & ~ & 6618   &   53877.6   &    51.0 & high/soft$^{l}$ \\
\hline
Cygnus X-1    & 5.6$^{e}$  & 2.5$^f$  &09.7 Iab$^{e}$ &   3724  &    52485.7  & 26.4 & high/soft$^{m}$\\
     ~        & ~  & ~ &  ~ &  3815  &    52702.7   &    55.9 & low/hard$^{n}$ \\
\hline
\hline
4U 1820$-$30    & 0.007$^{g}$ & 7.6$^h$  & low-mass$^{g}$ &   6633  &    53959.9   &    25.2 & BB$^{p}$ \\
       ~        & ~ & ~ & ~ & 6634  &    54028.2   &    25.1 & BB$^{p}$\\
       ~        & ~ & ~ & ~ & 7032 &     54044.6   &    46.3 & BB$^{l}$\\
\hline
Cygnus X-2     & 9.843$^{i}$ & 7.2$^j$ & A5-F2$^{i}$ &  8170  &    54337.7  & 78.7 & NB, HB, FB$^{q}$ \\
     ~         & ~ & ~ &  ~ &  8599  &    54335.2   &    70.6 & NB, HB, FB$^{q}$ \\
\hline
\hline
\end{tabular}
\vspace{-0.2in}
\tablecomments{The table above lists the name, orbital period, and the
companion type (where known) for the X-ray binaries included in our
sample.  The {\it Chandra} observation identification number,
observation start time, and exposure time is also given for each
observation.  Finally, the ``state'' in which the source was observed
is provided.  The first three sources are black hole candidates, the
last two sources are low-mass X-ray binaries harboring neutron
stars.\\
$^{a}$ Hynes et al.\ (2003).
$^{b}$ Zdziarski et al.\ (2004).
$^{c}$ Torres et al.\ (2006).
$^{d}$ Based on its coordinated near to the Galactic center.
$^{e}$ Gies \& Bolton (1982).
$^{f}$ Bregman et al.\ (1983).
$^{g}$ Stella, Priedhorsky, \& White (1987).
$^{h}$ Kuulkers et al.\ (2003).
$^{i}$ Cowley, Crampton, \& Hutchings (1979).
$^{j}$ Kuulkers et al.\ (1999).
$^{k}$ Miller et al.\ (2004).
$^{l}$ This work.
$^{m}$ Feng, Tennant, \& Zhang (2003).
$^{n}$ Miller et al.\ (2006a).
$^{p}$ ``Banana'' branch; Cackett et al.\ (2009).
$^{q}$ Normal branch, horizontal branch, flaring branch; Schulz et al.\ (2009).
}
\end{footnotesize}
\end{center}
\end{table}

\clearpage

\centerline{~\psfig{file=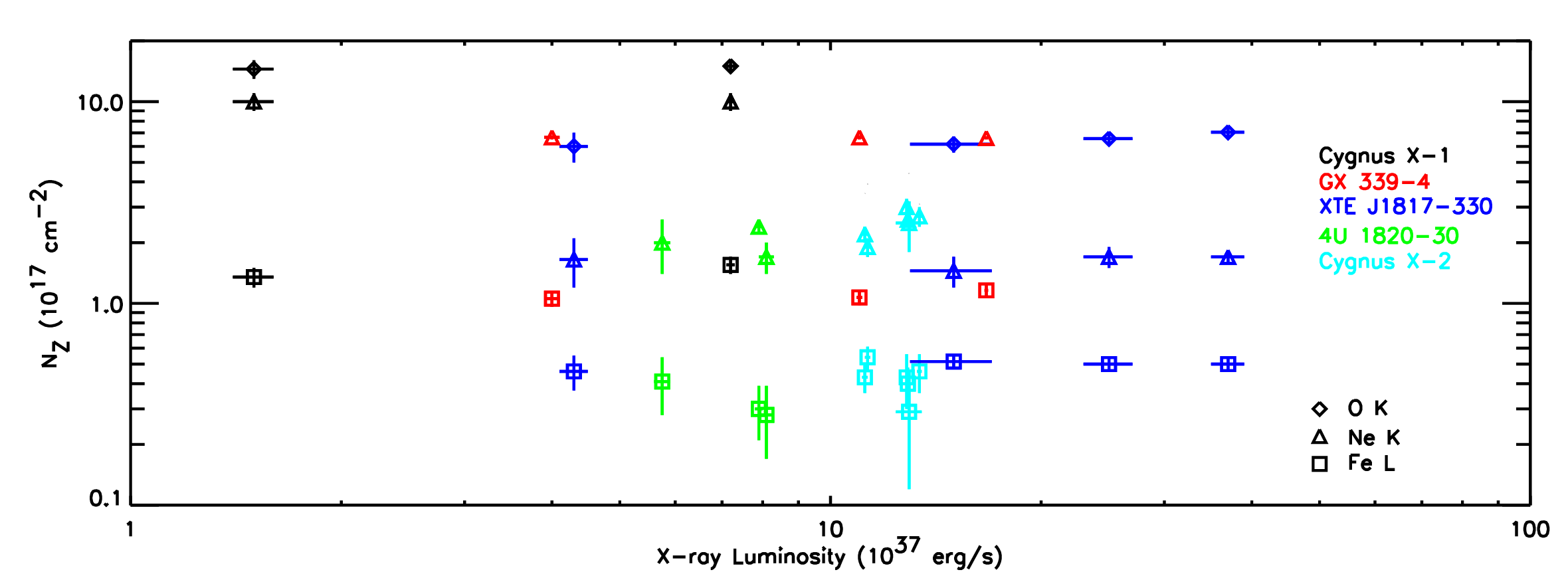,width=6.5in}~}
\figcaption[t]{\footnotesize The plot above shows the column density
measured in specific low energy absorption edges versus luminosity,
for black hole and neutron star X-ray binaries.  Errors on column
density are $1\sigma$ errors.}
\medskip


\end{document}